\documentstyle[12pt,epsf]{article}
\input epsf

\def\L{{\cal L}}
\def\O{{\cal O}}
\def\H{{\cal H}}
\def\li2{{\rm Li}_2}

\def\gev{{\rm GeV}}

\def\roughly#1{\,\,\raise.3ex\hbox{$#1$\kern-.75em\lower1ex\hbox{$\sim$}}\,\,}

\def\beq{\begin{equation}}
\def\eeq{\end{equation}}
\def\bea{\begin{eqnarray}}
\def\eea{\end{eqnarray}}

\def\Mpl{M_{\rm Pl}}

\def\fivecube{SU(5)_1\times SU(5)_2\times SU(5)_3}
\def\gws{SU(3)\times SU(2)\times U(1)}
\def\gwsp{SU(3)'\times SU(2)'\times U(1)'}
\def\one{{\bf 1}}
\def\two{{\bf 2}}
\def\three{{\bf 3}}
\def\five{{\bf 5}}
\def\fivebar{{\bf \bar 5}}
\def\Qb{\bar Q}
\def\bQ{\bar Q}
\def\qb{\bar q}
\def\labu{\bar\lambda_1^2}
\def\labd{\bar\lambda_2^2}
\def\labt{\bar\lambda_3^2}
\def\lab{\bar\lambda^2}
\def\det{{\rm det}}
\def\tr{{\rm Tr}}
\def\eightpisq{{1\over 8\pi^2}}
\def\fourpisq{{1\over 16\pi^2}}

\def\prl#1#2#3{Phys. Rev. Lett. {\bf #1} (#2) #3}
\def\mpl#1#2#3{Mod. Phys. Lett. {\bf A#1} (#2) #3}

\def\la#1{\lambda_{#1}}
\def\gsim{{~\raise.15em\hbox{$>$}\kern-.85em
          \lower.35em\hbox{$\sim$}~}}
\def\lsim{{~\raise.15em\hbox{$<$}\kern-.85em
          \lower.35em\hbox{$\sim$}~}}

\begin{document}
\begin{titlepage}
\begin{center}
May 1997  \hfill    CERN-TH/97-98\\
               \hfill    hep-ph/9705307
\vskip .2in
{\large \bf 

Dynamical Soft Terms with Unbroken Supersymmetry}
\vskip .3in

\vskip .3in
S. Dimopoulos$^{(a,b)}$, G. Dvali$^{(a)}$, R. Rattazzi$^{(a)}$, and 
G.F. Giudice$^{(a)}$\footnote{On leave
of absence from INFN, Sez. di Padova, Italy.} \\[.03in]

{$^{(a)}$\em Theory Division, CERN\\
     CH-1211 Geneva 23, Switzerland}

{$^{(b)}$ \em Physics Department, Stanford University\\ 
Stanford, CA 94305, USA}

\end{center}
\vskip .2in
\begin{abstract}
\medskip

We construct a class of  simple and calculable theories for the 
supersymmetry breaking soft terms. They are based on quantum modified 
moduli spaces.  These theories do not break supersymmetry in their 
ground state; instead we postulate that we live in a supersymmetry 
breaking plateau of false vacua. We demonstrate that tunneling from 
the plateau to the supersymmetric ground state is highly 
supressed. At one loop, the plateau develops a local minimum 
which can be anywhere between $10^8$ GeV and the grand unification 
scale. The value of this minimum is the mass of the messengers of 
supersymmetry breaking. Primordial element abundances  
indicate that the messengers' mass is smaller than $10^{12}$ GeV. 

\end{abstract}
\end{titlepage}

\section{Introduction}

Gauge mediated theories of supersymmetry breaking \cite{ancient}
have recently attracted a great deal of attention  
becuse they solve the supersymmetric flavor problem.
Beginning with the pioneering papers of Dine, Nelson and
collaborators \cite{DiNeSh}, a lot of effort has been devoted
to the construction of explicit realistic models with dynamical 
supersymmetry breaking \cite{HoIzYa}-\cite{YS}.
It seems fair to say that these models, when they work, are
quite involved. 
 
In this paper we wish to show how to construct simple theories
of the supersymmetry breaking soft terms. The key ingredient
that will allow us to do that is to relax the requirement that
supersymmetry is spontaneously broken in the full theory.
We instead will be content to postulate that we live in a 
supersymmetry breaking local minimum located on a plateau
of flat directions.

In section 2 we will present the general class of theories we 
consider. In section 3 we will demonstrate the it is not dangerous 
to live on a plateau: 
the tunneling amplitude into the supersymmetry preserving 
true vacuum is infinitesimally small.
In section 4  we will present explicit models implementing the ideas
of section 2. In section  5 we will show that light element abundances
place a limit that the messenger mass 
--which is related to  the position of the local minimum in 
field space--
has to be less than $10^{12}$ GeV. We conclude in section 6 with some
brief comparisons of our models with earlier work.

\section { The General Idea}
We are interested in theories based on a gauge group 
\beq
G=SU(5)_W\times G_B \times G_S.
\label{group}
\eeq
By $SU(5)_W$ we mean indeed $\gws$ as the dynamics
we describe takes place below the GUT scale. Nonetheless, as we want
to preserve perturbative gauge unification, we will only consider
messenger sectors involving complete $SU(5)_W$ multiplets.
 The role of the remaining factors,
as it will be clear below, is to provide a stable vacuum with
broken supersymmetry. We consider the following simple
form for the messenger sector superpotential
\beq
W=\lambda_1R q\qb +\lambda_2 R Q\Qb
\label{simple}
\eeq
where $R$ transforms non trivially only under $G_B$. The
$q\oplus \qb$ are vectorial under $SU(5)_W$, {\it i.e.} $\five\oplus
\fivebar$, but singlets of $G_S$. Finally, $Q\oplus \Qb$ are
a vectorial representation of $G_S$, but singlets of $SU(5)_W$.
All the fields transform non-trivially under $G_B$. 
Just to make contact with usual gauge mediated models, we anticipate
that the $q,\qb$ will be play the role of the messengers, the $Q,\Qb$
will provide a scale via their condensation, and $R$ will play the role 
of the usual $X$ field getting both a scalar and an $F$ component vacuum 
expectation value (VEV)  from the strong dynamics.

Furthermore we are interested in theories with the following two
properties at the classical level:
\begin{description}
\item [I)] $R$ contains just one $D$-flat direction, or equivalently
  there exist just one holomorphic invariant $u=u(R)\sim R^k$, which is
  a monomial made out of $R$ elements and $k$ is an integer. Along this flat
   direction the gauge factor $G_B$ is broken to a subgroup $H_B$ and
  all elements in $R$ but one, which we parametrize by $u$, are eaten
  by the SuperHiggs mechanism. For notational purposes
  we will in what follows parametrize the flat direction with a field
  $X=u^{1/k}\sim R$, since this is the one with canonical K\"ahler
  metric at large field values.

\item [II)] Along $u$ the $q,\qb$ and $Q,\Qb$ pair up to get
 masses which are respectively $\sim \lambda_1 R$ and $ \sim\lambda_2 R$.
\end{description}
Thus far away along $u\not = 0$ the low energy theory
has gauge group $G=SU(5)_W\times H_B\times G_S$ and the sole matter content
is given by the MSSM fields plus the gauge singlet $u$. The
additional group factors are just pure gauge with no matter.
Furthermore we are going to assume that the only sizeable
non perturbative low energy dynamics is the one associated with $G_S$.
On a case by case basis we will  check that $H_B$ is either too weak
or it does not affect the effective superpotential.
Above the scale of $G_S$ confinement the effective theory has
$W=0$.  The effective holomorphic scale $\Lambda_{eff}$ of the 
low energy $G_S$ theory is
given by the  1-loop matching of the  gauge couplings in the microscopic
and macroscopic theories at the scale $X$ \cite{russians}
\beq
\Lambda_{eff}^{3\mu_{G_S}}= \Lambda^{3\mu_{G_S}-\mu_Q}(\lambda_2 X)^{\mu_Q}
\label{matching}
\eeq
where $\mu_{G_S}$ and $\mu_Q$ are the Dynkin indices for the adjoint
and $Q\oplus \Qb$ representations of $G_S$. Below the scale $\Lambda_{eff}$
the $G_S$ charges are confined in massive states, and the only
effect of the $G_S$ dynamics on the low energy theory is to generate
a superpotential for $X$ via gaugino condensation \cite{gluino1,gluino2}
\beq
W_{eff}= <\lambda_S\lambda_S>= \Lambda_{eff}^3=\Lambda^{3-\mu_Q/\mu_{G_S}}(
\lambda_2 X)^{\mu_Q/\mu_{G_S}}.
\label{weff}
\eeq
In the limit in which the other gauge group factors are treated as
weak the above result is exact. It can indeed be established by using
holomorphy and an $R$-symmetry which has no $G_S$ anomaly \cite{seibergone}. 
Under this symmetry $X$ has charge $2 \mu_{G_S}/\mu_Q$, consistent with
the above $W_{eff}$ having charge 2.

The further requirement that we are going to make in order
to generally characterize our models is that $\mu_Q=\mu_{G_S}$. For
the class of theories that satisfy this identity $W_{eff}$ is linear in $X$
\beq
W_{eff}=\lambda_2\Lambda^2 X.
\label{linearw}
\eeq
An example of this possibility is given by $G_S=SU(N)$ with $Q+\Qb$
corresponding to $N$ flavors of fundamentals. This is the crucial
property that allows us to build simple and realistic theories with locally 
broken supersymmetry. Eq. (\ref{linearw}) corresponds
in fact to the simplest O'Raifertaigh model: $F_X=\partial_XW_{eff}=\lambda_2
\Lambda^2$, so that supersymmetry is broken at any point on the
$X$ complex line. For $\mu_Q\not =\mu_{G_S}$ the scalar
potential $|\partial _XW_{eff}|^2$ would push $X$ either to the origin
or to infinity, where supersymmetry is in general  restored.
Indeed, in models with the simple microscopic  superpotential in eq. \ref{simple},
at the point $X=0$ there are in general other flat directions
involving the massless $q$ and $Q$, where supersymmetry can be restored.
The existence of a plateau $V_{eff}\sim |\partial_X W_{eff}|^2\sim \lambda_2^2
\Lambda^4$ insures that $X$ may stabilize in some other way
far away from supersymmetry restoring points. This is the novelty of
the models we discuss: there is no need to break supersymmetry in
the true ground state.

The result in eq. (\ref{linearw}) can also be derived by using
the superpotential from confinement derived in Ref. \cite{quantumc} 
(see also Ref. \cite{veneziano})
and by integrating out the massive composite  superfields.

Up to this point we have devised a general class of models for
which there exists a direction $X$ in field space along which
$W_{eff}\propto X$. In such a situation the curvature of the scalar
potential is completely determined by the K\"ahler metric. We are interested
in the region $\lambda_2 X\gg \Lambda$, where it is consistent to consider only
the perturbative corrections to the K\"ahler potential. These we know
how to calculate.
The tree level K\"ahler potential is given by (see {\it eg.} Ref. 
\cite{popprand}) 
$K_{tree}= (u u^\dagger)^{1/k}=X X^\dagger$. The radiative corrections
are due to loops involving the heavy superfields $Q,\Qb$ and $q,\qb$ 
and the heavy vector superfields in the coset $G_B/H_B$. All
these particles  get a mass from the $X$ VEV. At the 1-loop
level the result has
the form
\beq
K(X,X^\dagger)= XX^\dagger \left [ 1 +\fourpisq\bigl ( C_Bg_B^2
-C_1 \lambda_1^2-C_2\lambda_2^2\bigr )\ln(XX^\dagger/M_{Pl}^2)+\dots \right ]
\label{oneloop}
\eeq
where the $C$'s are positive Casimirs and the coefficient of the logarithm
just corresponds to the 1-loop anomalous dimension of the original field
$R$. For large
logarithms the higher loop terms can be resummed via the Renormalization
Group (RG) so that in the leading log approximation the K\"ahler metric
reads
\beq
K(X,X^\dagger)= XX^\dagger Z_R\left (XX^\dagger/M_{Pl}^2\right ).
\label{leadlog}
\eeq
In the above, $Z_R$ represents the $R$ wave function renormalization
and $X$ the bare field.
Then the effective potential is just given by 
\beq
V_{eff}= {|\partial_X W|^2\over \partial_X\partial_{X^\dagger} K}\simeq
{\lambda_2^2 \Lambda^4\over  Z_R(XX^\dagger)}
\label{effpot}
\eeq
where  we have approximated  
$\partial_X\partial_{X^\dagger} K$ with $Z(XX^\dagger)$, neglecting
small finite corrections. The logarithmic evolution of $Z_R$ with
$|X|$ can generate local minima in the effective potential. Indeed
the potential energy behaves qualitatively like a squared effective Yukawa 
coupling. As it can be  inferred already from eq.~(\ref{oneloop}), in
the limit in which $g_B\ll \la{1},\la{2}$ we have that $1/Z_R$
grows at large $X$ thereby preventing a runaway behaviour. Indeed
one can see this as a signal of the fact that pure Yukawa theories
become strong in the ultraviolet by developing a Landau pole. On
the other hand, in the opposite limit of $g_B\gg \la{1},\la{2}$ 
the potential grows at small $X$. This can prevent $X$ from reaching the origin
or other branches of moduli space where supersymmetry may be restored.
 It is thus clear that the competition
between these two effects, which arises when $\lambda_{1,2}\sim g_{B}$
may stabilize the potential at some local minimum $X_0$. Notice that
it is crucial that $X$ comes form a field $R$ which is charged
under some gauge group $G_B$. This is precisely what provides
the gauge contribution $g_B^2$ without which $X$ is always  pushed 
towards the origin, where we loose control of our approximation (and
in most cases restore supersymmetry).
 The location
of this $X_0$ is determined by the stationary point of the RG equation
for $Z_R$
\beq
8 \pi^2 {d \ln Z_R\over d\ln|X|}=C_B g_B^2(|X|)-C_1{\lambda_1^2\over
 Z_RZ_qZ_{\qb}}
-C_2{\lambda_2^2\over Z_RZ_QZ_{\Qb}}=0
\label{stationary}
\eeq
where $g_B$ and the $Z$'s are respectively the running gauge coupling and
 wave functions  (of the various fields) evaluated at the scale $X$,
while $\lambda_{1,2}$ are the bare Yukawa couplings, {\it i.e.} 
renormalized at the cut-off scale $M_{Pl}$.  Then the  expressions 
multiplying $C_1$ and $C_2$ represent
respectively the effective running Yukawa couplings $\bar \lambda_1^2(X)$ 
and $\bar \lambda_2^2(X)$.
 This is just the usual
Coleman-Weinberg mechanism. We have a ``perturbative'' dimensional
transmutation, by which it is consistent to expect the minimum $X_0$
to be in the range $\Lambda \ll X_0\ll \Mpl$. As we will explain
in detail in the next sections, this is precisely the range
which is interesting for building realistic models.
It is also the range for which the minimum on the plateau,
even if it is not the true ground state, is nonetheless
stable on cosmological time scales (see the next section).
Finally, to be more precise, in order to insure that the stationary
point be a minimum one should verify that $d^2 Z_R/d (\ln X)^2$ be negative.
This equation, see eq.~(\ref{stationary}), involves both the gauge
$\beta$-function and the $\gamma$'s of the other fields, and has
to be checked case by case.

Before going further we would like to remark
that this mechanism for stabilizing a tree level flat potential
in a supersymmetric theory is just Witten's inverse hierarchy 
\cite{witten}. The amusing and new thing is that in our case
the superpotential is generated non-perturbatively rather 
than being an input at tree level. In other words we have a {\it double} 
dimensional 
transmutation
\begin{description}
\item [1)] $<F_X>=\lambda_2\Lambda^2$ is provided by gaugino condensation.
\item [2]) $<X>$ is stabilized at some point by the Coleman-Weinberg 
mechanism. 
\end{description}
It is interesting that in the region
$X\gg \Lambda$ we can determine the exact superpotential
together with a fairly accurate perturbative approximation to the
K\"ahler potential. This allows us to safely control the
full scalar lagrangian. We stress once more that, as in Witten's
model, it is crucial that $X$ be part of a charged multiplet
in order to have balancing gauge and Yukawa contributions in $V_{eff}$.
The issue of how a non-perturbative superpotential is stabilized
by RG evolution is a subtle one, and our result disagrees with 
ref.~\cite{shirman}. The conclusion of that
paper is that $V_{eff}\sim \lambda^2\Lambda^4/Z_RZ_QZ_{\Qb}$,
{\it i.e.} proportional to the running Yukawa coupling, rather than
just $V_{eff}\sim \lambda_2^2\Lambda^4/Z_R$ which we believe to be the correct
result. Notice that the result of ref.~\cite{shirman} would allow to stabilize
the potential even for a gauge singlet $R$, as $Z_QZ_{\Qb}$ already
involve a gauge coupling contribution with sign opposite to
that of Yukawa's. It seems to us that this result 
violates holomorphy. We believe that the correct approach is to integrate 
the RG  without rescaling the chiral superfields. If the
fields are rescaled,  a spurious  dependence on $  \lambda^\dagger$ and
$X^\dagger$ is induced both
in the running yukawa coupling and in $\Lambda_{eff}$, the latter via
the scale anomaly \cite{russians}.
%In other words, by performing a non-holomorphic rescaling of the chiral
%superfields also the matching equation (\ref{matching}) gets 
%$X^\dagger$ dependent corrections. 
In the end, however, this spurious dependence should cancel out and the
result for $W_{eff}(X)$ be  consistent with holomorphy. 
We use a scheme in which the 
fields are not rescaled, so that holomorphy is manifest in each step.

What is the phenomenology of these models?
At the decoupling of the messengers $q,\qb$, effective operators
involving the MSSM superfields and $X$ are generated. These
are the sources of sparticle masses after $X$ gets an $F$-component
VEV. The evaluation of the relevant diagrams
in component has been performed in ref.~\cite{diag}.
The gaugino masses just correspond to the 1-loop contribution of
the messengers to the gauge $\beta$-function:
\beq
\L_{\tilde g_i} =\int d^2\theta {N}{\alpha_i\over 4 \pi}\ln (\lambda X) 
W_\alpha  W^\alpha
\label{gaugeop}
\eeq
where $N$ is the number of messenger flavors and where $i=1,2,3$ 
runs over the $\gws$ couplings. The resulting gaugino
masses are thus
\beq
m_{\tilde g_i}=N{\alpha_i\over 4 \pi}{F_X\over  X}=N{\alpha_i\over 4
 \pi}\left ( {\lambda_2 \Lambda^2\over Z_X X}\right ).
\label{gaugemass}
\eeq
where $Z_X$ indicates the wave function $Z_R(XX^\dagger)$ at the minimum.
Phenomenology requires $\lambda\Lambda^2/Z_X X\sim 10-100$ TeV.
The sfermion masses arise from 2-loop diagrams, and they are also related to
the $\beta$ function coefficients that govern the evolution of
the chiral superfield wave function. For instance the squarks get a 
contribution $\propto \alpha_3^2$ and given by
\beq
\L_{\tilde Q}=\int d^4\theta N {8\over 3}({\alpha_3\over 4 \pi})^2\left (
{1\over 2}\ln^2(XX^\dagger/\mu_{UV}^2)+\ln(XX^\dagger/\mu_{UV}^2)
\ln(\mu_{IR}^2/XX^\dagger)\right ) QQ^\dagger
\label{squarkop}
\eeq
where $\mu_{UV}$ and $\mu_{IR}$ are respectively the
ultraviolet and infrared cut-off of the diagram. The above expression
has a very simple interpretation. The $\ln^2(\mu_{UV})$ arises 
from the integration region where both loop momenta are $\gg X$,
while the infrared divergent piece is determined by momenta $<X$
in the external loop. This explains the relative $1/2$
factor. By expanding the $F$ components of $X$ the above expression
contributes both to the chiral preserving squark masses and
to the $A$-terms. The latter are generated by the simple powers of $F_X$
after eliminating the auxilliary quark $F$-fields. Notice that the
contribution to the $A$-terms is of the form $\alpha_3^2\ln(X/\mu_{IR})$,
which is precisely the result one would get by solving the
1-loop RG equation below the messenger scale. Notice also that it
is precisely the relative $1/2$ coefficient amont the two $\ln^2$ contributions
in the above equation that ensures that there be no ultraviolet 
divergent contribution to $A$-terms. This is indeed a constraint one can use
 to derive eq.~(\ref{squarkop}). Finally we report the expression
for sfermion masses
\beq
 \tilde m^2=2N \left ({\lambda_2\Lambda^2\over Z_X X}\right )^2\sum_{i=1}^{3}
k_i({\alpha_i\over 4\pi})^2
\label{smasses}
\eeq
where the sum is over the $\gws$ factors. Moreover $k_1=(3/5)(Y/2)^2$ with 
hypercharge normalized such that $Q=T_3+Y/2$, $k_2=3/4$ for $SU(2)$ doublets
and zero for singlets while $k_3=4/3$ for color triplets and zero otherwise.

Since $X$ and $F_X$ are independent scales, we have that their
relative ratio can be wherever it is consistent with phenomenology and
with our approximations.
In particular, and we will discuss this in more detail later on,
the messenger scale $X$ can be anywhere between $10^8$ and $10^{14}$ GeV.
Consequently  the number of messenger flavors can be bigger than the minimum
$\sim 4-5$ which one gets for $X\sim 10^5$ GeV  by requiring perturbative
gauge unification \cite{CaMu}. By varying $G_S$ we can explore
essentially all possible gauge mediated boundary conditions on
sparticle masses.

We have now to discuss why such a local mimimum on
a plateau is cosmologically acceptable.

\section{The Stability of our Universe}

%%%%%%%%%%%%%%%%
% FIGURE 0
%%%%%%%%%%%%%%%%
\begin{figure}[t]
%   \vspace{-4cm}
   \epsfysize= 8.5cm
   \epsfxsize=10.2cm
   \centerline{\epsffile{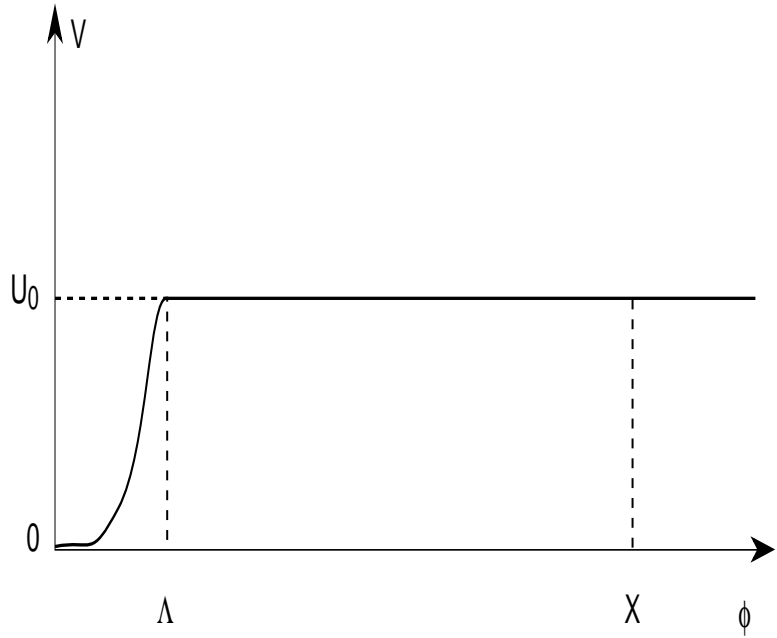}}
%   \vspace*{-15cm}
\caption{\label{plateau} 
An idealized flat potential with $V(\phi>\Lambda)=U_0\sim \Lambda^4$ and
$V(0)=0$.
}
\end{figure}
%%%%%%%%%%%%%%%%

The potential $V_{eff}(X)$ in our class of models can
be pictured as a very flat (with logarithmic curvature) plateau
at $X\gg \Lambda$. Moreover we know that classically at $X=0$ there are other 
branches
of moduli space. Some of these are supersymmetry
preserving even at the quantum level.
We expect that the transition between the plateau and these other directions
of lower energy takes place smoothly in the region $X\lsim \Lambda$. 
Physically what happens
in this region, is that some mesons or baryons of $G_S$ become tachionic
and develop a VEV. In the Appendix we discuss this in some
more detail for a specific model.
In this section we are going to discuss the quantum stability
of a false vacuum with these properties. In order to get an estimate
of the rate of tunneling we can reduce to a one dimensional potential 
$V(\phi)$.
In some of the cases of interest, the region 
$\phi< \Lambda$ corresponds to more than one vacuum or flat direction.
Still our conclusions are not  qualitatively modified
in the realistic case. 
%Why it is so will be more clear at the end of this
%discussion. 
Furthermore in order to make the discussion 
simpler, we will study the rate of true vacuum bubble nucleation for a
truly flat plateau, in
the limit in which $\Lambda\ll X$. This is shown in Fig. 1.
Indeed, at first sight, this may look like a bad approximation as it 
eliminates the potential barrier that separates the physically interesting,
but false vacuum, from the supersymmetry preserving one. But this is too naive
since in field theory the potential energy $\H$ consists of both a gradient
and a static piece
\beq
\H=\int d^3 x\left \{ (\nabla \phi)^2 + V(\phi)\right \}.
\label{hamiltonian}
\eeq
When the potential $V(\phi)$ is flat, it is the gradient energy $(\nabla\phi
)^2$ that provides the barrier in the functional  space $\{ \phi\}$.
Consider, indeed, a homogeneous field configuration where $\phi(x)\equiv
X$ over all space. Then for any {\it small} and {\it local} variation
$\delta \phi(x)$ of the background field, the static energy is unchanged
while the gradient contribution is positive definite
\beq
\Delta \H=\Delta \H_{grad}+\Delta \H_{stat}=\Delta \H_{grad}> 0.
\label{deltaH}
\eeq
It is also clear that the larger $X/\Lambda$ the more gradient
energy must be generated in order to reach lower potential energy,
so the higher the barrier in field space is. We thus expect that
for large $X/\Lambda$ the rate of quantum tunneling into the true vacuum will
be suppressed. The rate of bubble nucleation can be then evaluated in
 the semiclassical approximation. This amounts to finding the action
for the bounce,
a spherically symmetric solution to the Euclidean equations of motion 
\cite{coleman}. In  terms of the bounce action $S_B$, the rate of
bubble formation per unit volume $\Gamma/V$ is just given by
\beq
{\Gamma\over V}=\mu^4 e^{-S_B}
\label{nucleation}
\eeq
where $\mu$ is a mass parameter determined by the calculation
of quantum fluctuations around the bounce and which we expect to
be of the order of the largest scale in the problem, $X$.
The bounce solution $\phi(r)$ satisfies the following 
boundary conditions and equations of motion
\bea
\dot \phi(0)&=&\dot \phi(\infty)=0\quad \quad\phi(\infty)=X\\
\ddot \phi&+&{3\over r}\dot \phi=\partial _\phi V(\phi).\\
\label{eqbounce}
\eea
The equation of motion describes a classical motion in a one dimensional
potential $=-V(\phi)$, with time variable $=r$
 and with a friction term whose strength decreases as $1/r$. The particle 
analogy is very useful and
one can apply  considerations similar to those made in Ref. 
\cite{coleman} for the thin wall bubble. Similar calculations
have been done in Refs. \cite{eweinberg,kusenko}.
 The size of the
bubble is simply  established by energy conservation. Consider indeed the 
motion in the plateau region $\phi\geq \Lambda$, see Fig. 1. Here $\phi$ is 
just a
massless field and the solution to eq.~(\ref{eqbounce}) is simply
\beq 
\phi(r)=(X-\Lambda)\left [1-\Bigl({R\over r}\Bigr )^2\right ]+\Lambda.
\label{wall}
\eeq
Notice that $\phi(\infty)=X$ and that the radius $R$ is an integration 
constant characterizing the size of the solution: $\phi(R)=\Lambda$. The 
kinetic energy $\dot \phi^2/2$ at $r=R$ is given by $(X-\Lambda)^2/2R^2\sim
X^2/2R^2$. In view of the presence of a dissipative force,
 this has to be smaller than the maximum
available energy $U_0$. Then it must be $R>X/{\sqrt 2 U_0}\gg 1/\Lambda$,
 much bigger than the typical scale of the potential $V(\phi)$. This large 
radius has two implications: i) as $X/\Lambda\to \infty$ the starting point
$\phi(0)\to 0$, where $\partial V=0$, since for $\phi(0)>0$, where $\partial V
\sim \Lambda^3$, the radius would be too small ${\cal O}(1/\Lambda)$, the only
scale in the potential; ii)  in the region $\phi<\Lambda$ the  friction
term are ${\cal O}(1/\Lambda R)$ with respect to the leading terms,
so that energy conservation holds up to order $1/R\Lambda\lsim \Lambda/X$.
Then from i), ii) and by equating potential and kinetic energy respectively
at $\phi\simeq 0$ and $\phi=\Lambda$ we find the bubble radius
\beq
U_0={X^2\over 2 R^2}\left (1+{\cal O}(\Lambda/X)\right )\quad\Longrightarrow
\quad R\simeq {X\over {\sqrt 2 U_0}}
\label{radius}
\eeq
More qualitatively we can describe the bounce solution as follows.
$\phi$ starts very close to the top, and there it stays until friction  is
small enough to allow the motion to $X\gg \Lambda$. (Remember that in
the particle analogy the potential is $-V(\phi)$ of Fig. 1.)
 At $r\sim R=X/
{\sqrt {2U_0}}$
the particle goes down the hill. This motion is determined by the
fundamental scale $\Lambda$ in $V$ and happens within a thin shell (short time)
$\Delta r\sim 1/\Lambda\ll R$. Its contribution to the action is
consequently negligible.  
Indeed we can write the action $S(R)$ as
\beq
S(R)= -{1\over 2}\pi^2 U_0 R^4 + 2 \pi^2 X^2R^2+ {\cal O}(\Lambda^3 R^3).
\label{actionR}
\eeq
The first term is from the bulk of the bubble $0<r\lsim R$ where 
$\phi\sim 0$. The
second comes from the massless region $r>R$, where
$\phi>\Lambda$, and by using eq.~(\ref{wall}). Finally, the third term
is an estimate of the contribution from the transition region, the thin shell.
 By
extremizing with respect to $R$ we find the same bounce radius
as eq.~(\ref{radius}). Neglecting the subleading term we then get the bounce
action
\beq
S_B=2\pi^2 {X^4\over U_0}\sim 2\pi^2
\left ({X\over \Lambda}\right )^2.
\label{bounce}
\eeq

The consistency of this approximation can also be verified by
noticing that eq.~(\ref{actionR}) is consistent with the virial theorem
$T+2U=0$, where $T,U$ are respectively kinetic and potential energy.
This is because the contribution from $r\simeq R$  is all potential while
$r>R$ is all kinetic.

% If the starting point $\phi(0)$ 
%is taken in the region
%where $\partial _\phi V$ is sizeable, {\it i.e.} between 0 and $\sim \Lambda$,
%then the particle will loose a lot of energy at small times due to the
%large friction term and will never reach $X\gg \Lambda$.  The solution
%of the equation of motion can then be patched into three regions
%\begin{description}
%\item [a)] $0<r<R \quad\quad \phi(r)\simeq 0$
%
%\item [b)] $R<r<R_1.\quad$ $\phi$ goes down the hill and $\phi(R_1)=\Lambda$.
%
%\item [c)] $R_1<r$. The field $\phi$ satisfies the
%massless equation of motion
%\beq 
%\phi(r)=X\left [1-\bigl({R_1\over r}\bigr )^2\right ]+\Lambda
%\label{wall}
%\eeq
%which goes asymptotically to $X$.
%\end{description}
%Finding the solution satisfying the
%above b.c. amounts to finding the value $\phi(0)$ from where the particle 
%starts its motion.

%For fixed $\Lambda$, the larger $X$ the bigger the bubble radius
%has to be, in order to avoid friction in the particle analogy.
%However when $R\gg 1/\Lambda$, the friction term can be neglected 
%in the evolution in region b), as the only scale in the potential
%is $\sim \Lambda$. This leads us to quickly conclude that
%^the motion in region b is very quick and that $R_1-R\sim 1/\Lambda$
%an be approximated with zero.  In this approximation the contribution
%to the action comes just from a) and c)

Finally with the calculated action we can estimate the lifetime of
our universe by using eq.~(\ref{nucleation}). Using the volume of the
past light cone we get consistency if $S_B> 500$. This is
easily satisfied, already for $X/\Lambda\sim 10$.
 
Having established the stability of the vacuum on
the plateau at the present, one may
ask the same question for the previous stages
of evolution of the universe. We do that by ``going backward
in time'', as later stages of cosmological evolution
are obviously more under control. So the first question is:
what are the conditions for $X$  not to be 
driven away by thermal corrections?
 Intuitively one might expect that at high temperature $T$ a
flat direction $X$ with renormalizable
 interactions is driven to the origin by the positive $\sim T^2$  
corrections 
 to its mass. However, this is true only for 
  $X \lsim T$, as for $X\gg T$  all the modes coupled to $X$
get heavy  and decouple. In these regions the potential is
determined by $1 /X$ suppressed couplings to the remaining light  
fields in the K\"ahler function.
This corrections provide a $\sim T^2/X$
curvature to the flat direction field. This is negligibly small if  
the maximum temperature of the universe, namely the  
reheating temperature, is much smaller than the X.
 For a reheat temperature
$T_{reheat}^3 < \Lambda^2 X$ 
gaugino condensation in the strong
group is not affected by the temperature corrections and the zero
temperature potential is dominant. Note that for values of $X_{today}$
 favoured   by our 
 models the above condition is {\it automatically}
satisfied when $T_{reheat}$ satisfies the
gravitino regeneration bound\cite{murayamacosmology}.
We conclude that it is consistent to survive in this local minimum
during the big bang. The next step behind deals with
the evolution of $X$ during inflation. Here, however,
the theoretical situation becomes less under control, especially 
in the absence of a well defined model of inflation.

% Still something can be said.
%In general during this stage the $X$ potential  receives a mass
%term $\sim H^2 |X|^2$, where $H$ is the Hubble parameter \cite{ddrt}.
%This can displace $X$ from its present value. This suggests that
%at least for 
%\begin{equation}
% m_X^2 \sim \left(\eightpisq\right)^2 {\Lambda^4 \over X^2_{today}} >
% H^2.
%\label{Hinf}
%\end{equation}
%the present day minimum is stable during inflation. Finally, for
%the pre-inflation epoch, one 

\section{Models}

We now illustrate some explicit models built along the 
ideas of Section 1. We name each of them by its $G_S\times G_B$
gauge group factor.

\vskip 0.5truecm
\noindent {\bf 4.1}\hskip 0.3truecm  ${\bf SU(2)\times SU(2)_{B1}
\times SU(2)_{B2}}$.
\vskip 0.3truecm

The simplest choice for the strong group is 
$SU(2)$ with a matter content given by four doublets
(two flavors) $Q_I,I = 1,\dots,4$. In the massless limit this sector has
a global $SU(4)$ flavor symmetry. Any anomaly free subgroup of it may be
weakly gauged and play the role of the balancing group $G_B$.
%In above examples, the balancing group $G_B$ was not asymptotically free,
%as it contained many flavors. On the other hand since $R$ must transform
%under it and at the same time couple to $Q,\bar Q$ and $q,\bar q$
%this conclusion seems difficult to avoid. One rather straightforward way is
%to increase both strong $G_S$ and $G_B$ groups. We briefly come back to this
%possibility later. 
However we find that the simplest way to realize the requirements
of section 1 is to consider
$G_B =SU(2)\times SU(2)\sim SO(4)\subset SU(4)$. For instance, the smaller 
subgroups  $U(1)$ or $SU(2)$ would not be asymptotically free. The need
for asymptotically free balancing group seems to be generic for the
mechanism to work.

The matter content under $SU(2)_S\times SU(2)_{B1}\times  
SU(2)_{B2}\times SU(5)$ is
\bea
Q = (2,2,1,1),\quad\bar Q = (2,1,2,1),\quad R = (1,2,2,1),\nonumber\\
  q_3\oplus q_2\oplus q_1= (1,1,2, 3\oplus 2\oplus 1)\quad\bar q_3\oplus
\bar q_2\oplus \bar q_1 = (1,2,1,\bar 3\oplus \bar 2\oplus \bar 1)
\label{2cubematter}
\eea
where the $q_i$'s have been split into their $\gws$ irreps. Notice that
we add a pair of standard model singlets $q_1,\bar q_1$  in order
to cancel the $SU(2)_{B1}\times SU(2)_{B2}$ global anomaly.
The tree superpotential is given by
\beq
      W =  R\left (\lambda Q\Qb +\lambda_3 q_3 \qb_3+ \lambda_2 q_2 \qb_2
+\lambda_1 q_1 \qb_1\right ).
\label{microw}
\eeq
The classically flat direction of interest is
$X = (det R)^{{1 \over 2}}$. Along $X$, we have
\beq
<R>=X\left (\begin{array}{cc}1&  0\\
                             0&1
                              \end{array} \right )
\label{rvev}
\eeq
so that $SU(2)_{B1}\times SU(2)_{B2}$ is
broken down to the diagonal $SU(2)_D$, and conditions {\bf I)} and {\bf II)}
of Sect. 2 are satisfied. In the regime in which $SU(2)_D$
(and the stardard model gauge group) is weak, the only relevant
non-perturbative dynamics is that of $SU(2)_S$. (We will later discusss how 
good  this assumption is.)  Then,
below the scale of $SU(2)_S$ confinement,  
the effective superpotential  is $W_{eff}=\lambda X \Lambda^2$. 
The minima of the effective superpotential are determined
by the RG evolution of the wave function, as discussed in sect.~2.
Let us analyze  this model in more detail. 
Indicating by $\mu$ the renormalization scale
we write the RG equations (RGE) in terms of $\tau =\eightpisq \ln(\mu/\Mpl)$. 
Using a vector notation, the gauge $\beta$ functions above the scale $X$
are
\beq
{d \over d \tau}({1\over g_B^2},{1\over g_S^2}, {1\over g_3^2},
{1\over g_2^2}, {1\over g_1^2})=(1,4,1,-3,-\frac{43}{5}).
\label{gaugerg2}
\eeq
where we have taken $g_{B1}=g_{B2}=g_B$, so that the balancing group is 
$SO(4)$. The equations for $Z_X$ and for the Yukawa couplings are
\bea
{d \ln (Z_X)\over d \tau}&=& 3g_B^2-2\lab -3\labt -2\labd -\labu  
\label{yukawarg2}\\
{d \ln \lab\over d \tau}&=& 6\lab + 3\labt +2 \labd +\labu -6g_B^2-3g_S^2 \\
{d \ln \labt\over d \tau}&=& 2\lab + 7\labt +2 \labd +\labu -6g_B^2-{16\over 3}
g_3^2 -\frac{4}{15}g_1^2\\
{d \ln \labd\over d \tau}&=& 2\lab + 3\labt +6 \labd +\labu -6g_B^2-3g_2^2 
-\frac{3}{5}g_1^2\\
{d \ln \labu\over d \tau}&=& 2\lab + 3\labt +2 \labd +5 \labu -6g_B^2\\
\eea
where $\bar \lambda_i$'s are the running couplings ({\it e.g.} $\lab=\lambda^2/
(Z_XZ_QZ_{\bar Q})$). 
By identifying $\mu$ with $X$, the minima of the potential
are given by the solutions to $d Z_X/d \tau=0$ with 
$d^2 Z_X/d\tau^2<0$. The quantity $\ddot 
Z_X$ is readily computed from eq.~(\ref{yukawarg2}), with the help
of the other RGE. Notice to this effect that the only terms that
can make $\ddot Z_X>0$ are the gauge contributions to the Yukawa
RGE. These contributions are indeed big for $\lab$, and they are 
sizeable also for $\labd$ and $\labt$. Therefore for
small $\labu$, the condition $\ddot Z_X<0$ cannot be satisfied.
On the other hand, when $\labu$ is bigger than the other Yukawa couplings,
one gets $\ddot Z_X<0$. Still, in this regime, at a given $X$,
$\labu$ cannot be made too big, otherwise it hits a Landau pole
below $M_{Pl}$. Such considerations determine the regions
of parameter space where our mechanism works. 
In fig.~1 we display, for different values of $\bar\lambda$,
the region in the plane $(g_B,\bar\lambda_1)$
for which there is a minimum at  $X=10^{10}$ GeV. We have assumed 
$\bar \lambda_3=\bar \lambda_2$ to reduce the number of parameters
and because $\bar \lambda_3\sim\bar \lambda_2$ is motivated
by GUT considerations. The boundaries of the region in the figure
are explained as follows: i) the right one is given by $\ddot Z_X<0$;
ii) the top one is essentially dictated by the absence of Landau poles
in $\labu$ below the Planck scale; iii) the left one
is determined by $\labu\leq 3 g_B^2$ from $\dot Z_X=0$, see 
eq.~(\ref{yukawarg2}). So it is clear that there is a significant region
of parameter space where this model behaves, in the observable MSSM
sector, as a conventional gauge mediated supersymmetry breaking model, with
messenger scale around $10^{10}$ GeV. We remark that 
this model shares similarity
with an interesting proposal by Izawa and Yanagida \cite{IzYa} and by
Intriligator and Thomas \cite{it},
where the role of our 
$R$ is essentially played by a set of gauge singlets. The phenomenological
version of that model,
mentioned in Ref. \cite{it} and discussed at length in Ref. \cite{HoIzYa},
is unfortunately not calculable. In that model,
like in ours, one must live in a local minimum, as the $Q\bar Q$ and the 
$q\bar q$
restore supersymmetry at $X=0$. However this local minimum, if it exists,
is presumambly at $X\lsim \Lambda$, as the Goldstino superfield
$X$ is a gauge singlet and the perturbative superpotential drives
it towards the origin  (the perturbative corrections to $K(XX^\dagger)$
were ignored in Ref. \cite{it,HoIzYa}). In this region 
the K\"ahler potential
is not calculable and one has just to assume: 1) the occurrence of such
a local minimum; 2) its quantum stability, as there is no obvious parametric 
suppression of the tunnelling rate at $X\sim \Lambda$.

%%%%%%%%%%%%%%%%
% FIGURE 1
%%%%%%%%%%%%%%%%
\begin{figure}[t]
%   \vspace{-4cm}
   \epsfysize= 8.5cm
   \epsfxsize=10.2cm
   \centerline{\epsffile{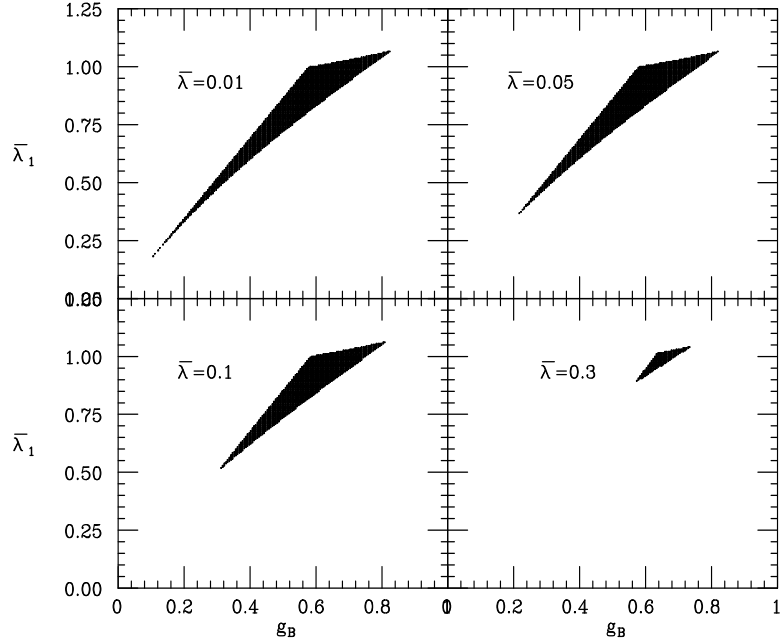}}
%   \vspace*{-15cm}
\caption{\label{su2cube} 
The region of parameters where the $SU(2)\times
SU(2)\times SU(2)$ model can have a local supersymmetry-breaking vacuum
at $\langle X\rangle =10^{10}$ GeV.
}
\end{figure}
%%%%%%%%%%%%%%%%

Around the minimum
the real part of $X$ gets a mass
\beq
m_X^2=\left ( \eightpisq\right)^2 \left ({\lambda \Lambda^2\over X}
\right )^2 {\ddot Z_X\over 8 Z_X^3}
\label{xmass2}
\eeq
which is parametrically similar to the sfermion masses in eq.~(\ref{smasses}).
At this stage,
the imaginary part of $X$ is the R-axion. Notice that $X$, the axion scale, 
as far as gauge mediation
is concerned, can consistently be in the $10^9-10^{12}$ GeV axion 
window \cite{axion}.
So $X$ also provides and interesting QCD axion candidate. However there is
still the possibility that ${\rm Im}X$ gets a mass from direct sources
of explicit $R$ breaking. One of these can arise from the mechanism that
cancels the cosmological constant \cite{bpr}. Another is the $SU(2)_D$
anomaly, but this is negligible when the balancing group is weak (see the 
comments
below).

Actually also $SU(2)_D$ has a low energy
dynamics that leads to an additional term $W_D\sim \Lambda_B
X^2$, where $\Lambda_B=\Lambda_{1,2}$ is the microscopic strong scale of
the balancing $SU(2)$ factors. The combination of the two gaugino condensates
restores supersymmetry, as the low energy $W\sim \Lambda^2X+\Lambda_B
X^2$ has no effective R-symmetry.
However, since we are interested in local minima we are not worried by 
that, as long as $\Lambda_B$ is
small enough that the supersymmetry 
preserving minimum is very distant. This is indeed
what happens for the values of $g_B(X)$ that are allowed  in Fig.~1. 
Notice that the relative shift in the vev is $\delta X/ X\sim
8 \pi ^2
\Lambda_B/m_X$, which for $g_B(X)<1$ is less than $10^{-20}$, while
another supersymmetry preserving minimum (in addition to
those at $X=0$) exists only at $X>M_{Pl}$! 

Some more comments are in order. These concern the range of $X$ where the
mechanism works. In particular $X$ cannot be too close to $\Lambda$,
otherwise our perturbative approximation to the K\"ahler potential
breaks down, due to non-perturbative corrections from the $SU(2)_S$
dynamics. Indeed at $X\gg \Lambda$ we expect the non-perturbative
corrections to $K(X,X^\dagger)$, to be dominated by the lowest
term in the operator product expansion.
 This  is given by $\int d^4 \theta (W_\alpha^2)(\bar
W_{\dot \alpha}^2)/(XX^\dagger)^2$, whose coefficient is $\sim \eightpisq$
as it is determined by a 1-loop diagram involving the heavy $Q,\bar Q$
and four external vector superfields. After substituting the gaugino
condensate the above operator leads to a correction
$\delta K(X,X^\dagger)\sim 8 \pi^2 \Lambda^4/(\lambda^2 XX^\dagger)$.
Thus the effects induced on the K\"ahler curvature vanish like 
$(\Lambda/X)^4$.
Using the above formula for $m_X^2$, we estimate that this 
non-perturbative effect can 
be reasonably neglected for $X\gsim 10^{9}$ GeV.

At the other end of the range, for large $X$, operators induced by Planck
scale phisics may affect
the flatness of the superpotential. We find that, barring the mass term 
$W= M_{Pl} X^2$, the lowest correction $X^4/M_{Pl}$ is tolerable when the
minimum is at $X\lsim 10^{11}$ GeV. For higher messenger scales this
term will also have to be suppressed. We notice that while the $X^2$ and $X^4$
operators are in principle allowed by the symmetries of the model,
it is not totally unreasonable to assume they are absent. This is
because in supersymmetric theories the absence of a given term in
an effective superpotential (now eq. (\ref{microw}) itself) does 
not necessarily have a symmetry
explanation in the low-energy theory. (This is for instance seen
in examples of strong gauge dynamics.) 

Before concluding we stress that the phenomenology
of this model is that of conventional gauge mediation with
two families of messengers in the fundamental of $SU(5)$.
One could consider generalizations based on
$Sp(m)\times Sp(m)\times SU(2m)$, where $m=1$ is just the original model.
These work essentially in the same way, but have $2m$ families of 
messengers.

\vskip 0.5truecm
\noindent {\bf 4.2}\hskip 0.3truecm  ${\bf SU(2n)\times SU(2n)}$ 
{\bf with adjoint matter.}
\vskip 0.3truecm
Another possibility  is to   consider 
 $G_S = SU(N)_S$ and $G_B = SU(N)_B$, with even $N=2n$ and 
with the
following matter content under $SU(N)_S\times SU(2)_B\times SU(5)$
\beq
Q = (N,\bar N,1),~~\bar Q = (\bar N,N,1),~~ R = (1,N^2-1,1),
~~ \bar q = (1,N,\bar 5), ~~q= (1,\bar N, 5).
\eeq
The classical superpotential is 
\beq
      W = \lambda R Q\Qb +\lambda_5 R q\qb+{k\over 3} R^3.
\label{class6}
\eeq
Again, at the classical level, there is a flat direction along which
only $R\not =0$
\beq
<R>=X\left (\begin{array}{cc}1&  0\\
                             0&-1
                              \end{array} \right )
\label{rvev6}
\eeq
where the entries represent $n\times n$ blocks. To motivate
this choice of model and $W$ we remind that
in the absence of a superpotential,
there exists an
$N-1$ dimensional family of flat directions parametrized by $N-1$  
invariants built from $R$: $X_p = (\tr R^p)^{1/p}$ with $p = 2...N$.
At a generic point on this vacuum manifold, $SU(N)_B$ is broken down to
$U(1)^{N-1}$ and there are just $N-1$ massless chiral and $N-1$ massless gauge
superfields. There are, however,
enhanced symmetry directions (lines) along which a non-Abelian
subgroup is unbroken. Here the number of massless gauge and chiral
superfields is bigger. A suitable superpotential can lift all but one
flat direction. For  $SU(mn)$ groups ($m,n$ integers) adding
$W={\rm Tr}R^{m+1}$, lifts all moduli but $u_m=\tr R^m$, along
which $SU(mn)\to SU(n)^m$. The special thing about $SU(2n)$ groups
is that this job is done by a trilinear invariant ${\rm Tr} R^3$.
This motivates our choice of even $N$ with $W$ given by eq.~(\ref{class6}). 

Now, along $X$, see eq.~(\ref{rvev6}), 
the balancing group is broken down to $SU(n)\times SU(n)$. All other
components of $R$ as well as
$Q,\Qb$ and $q,\qb$  have a mass of order $X$ and decouple.
The low energy dynamics of $SU(N)_S$ generates a linear superpotential
$W=\lambda X\Lambda^2$. One can go through the same analysis we
applied to $SU(2)^3$. In the models with $N=2n\leq 4$ the balancing
group is not asymptotically free, and this makes it impossible
to satisfy the minimum condition $\ddot Z_X<0$. Moreover models with
$N\geq 10$ have too many messengers, so that Landau poles
in the evolution of the SM gauge coupling are hit before the GUT
scale, unless $X$ is very close to $M_G$, bringing partially back
the supersymmetric flavor problem. We find that a model which works
very well over a wide range of scales is $SU(6)\times SU(6)$.
Let us discuss it in more detail.
 The gauge coupling RGE are
\beq
{d \over d \tau}({1\over g_B^2},{1\over g_S^2}, {1\over g_3^2},
{1\over g_2^2},{1\over g_1^2})=(1,12,-3,-7,-{63\over 5}).
\label{gaugerg6}
\eeq
The equations for $Z_X$ and for the Yukawa couplings are
\bea
{d \ln (Z_X)\over d \tau}&=& 12 g_B^2-6\lab -3\labt -2\labd -{16\over 3}
\bar k^2  \label{yukawarg6}
 \\
{d \ln \lab\over d \tau}&=& {53 \over 3}\lab + 3\labt +2 \labd +{16\over 3}
\bar k^2 -
{71\over 3}g_B^2-{35 \over 3}g_S^2 \\
{d \ln \labt\over d \tau}&=& 6\lab + {44\over 3}\labt +2 \labd +{16\over 3}
\bar k^2 -{71\over 3}g_B^2-{16\over 3} g_3^2 -\frac{4}{15}g_1^2\\
{d \ln \labd\over d \tau}&=& 6\lab + 3\labt +{41\over 3} \labd +{16\over 3}
\bar k^2 -{71\over 3}g_B^2 -3g_2^2 -\frac{3}{5}g_1^2\\
{d \ln \bar k^2\over d \tau}&=&-3 {d \ln (Z_X)\over d \tau}\\
\eea
where again the barred Yukawas indicate running couplings.

%%%%%%%%%%%%%%%%
% FIGURE 2
%%%%%%%%%%%%%%%%
\begin{figure}[t]
%   \vspace{-4cm}
   \epsfysize= 8.5cm
   \epsfxsize=10.2cm
   \centerline{\epsffile{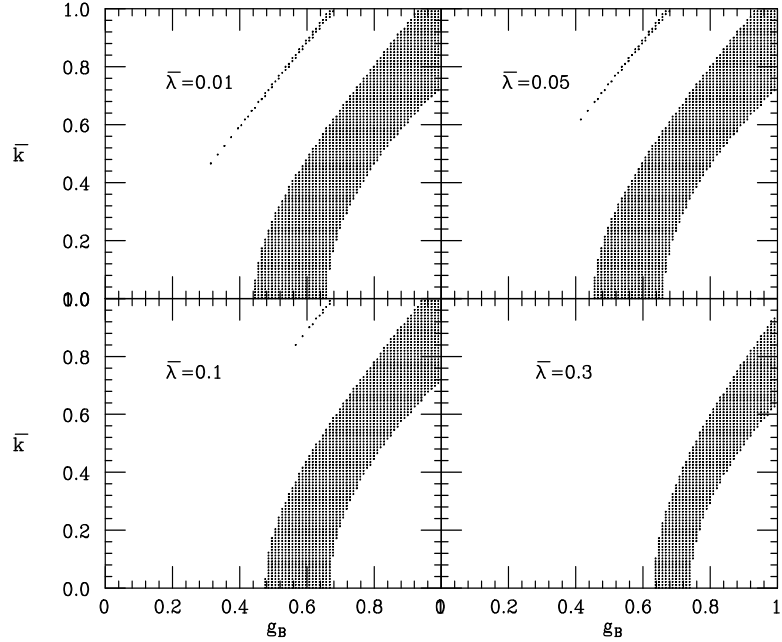}}
%   \vspace*{-15cm}
\caption{\label{su6square} 
The region of parameters where the $SU(6)\times
SU(6)$ model can have a local supersymmetry-breaking vacuum
at $\langle X\rangle =10^{10}$ GeV.
}
\end{figure}
%%%%%%%%%%%%%%%%

In Fig.~2 we show, for different values of $\bar \lambda$
 the region in the $(g_B,\bar k)$ plane where the model
leads to a stable minimum at $X=10^{10}$ GeV. Again we assumed $\bar\lambda_2
=\bar \lambda_3$. The two disjoint regions arise as follows.
The thin region around the line $g_B=(2/3)~\bar k$, corresponds
to the points where $\dot Z_X=0$ is saturated essentially by $\bar k$,
see eq.~(\ref{yukawarg6}). To the immediate right of the thin area, 
$\bar \lambda_{2,3}$ are small but
non zero. This is enough for the negative contribution of $g_3$
and $g_2$ to their RGE to make $\ddot Z_X$ positive.
This excludes this intermediate area. However when $\bar \lambda_{2,3}$
become large enough, the positive terms in their RGE dominate and $\ddot Z_X$
becomes negative again. This explains the big allowed band to the right. 
However
as $\bar \lambda_{2,3}$ are further increased, they hit a Landau pole below 
$M_{Pl}$, and the region of large $g_B$ is again excluded. The range
where the model works becomes even wider as the scale $X$ is increased
above $10^{10}$ GeV, as the constraint from Landau poles gets weaker. Also,
as for the previous model, the effects on $W$ of the low energy $SU(3)^2$ are
negligible as soon as $g_B(X)\lsim 1$.

\vskip 0.5truecm
\noindent {\bf 4.3}\hskip 0.3truecm  ${\bf SU(5)^3}$. {\bf A special model.}
\vskip 0.3truecm

As a special case of our general idea one can consider
the possibility of identifying $Q,\Qb$ with $q,\qb$. In other
words the messengers themselves trigger supersymmetry breaking. This
can be achieved in a model with full gauge group
$\fivecube$ and with matter content given by
$R(\one,\five,\fivebar)$, $Q(\fivebar,1,\five)$ and  
$\bQ(\five,\fivebar,\one)$.
For phenomenological applications we are interested
in a  variant (low-energy limit) of this model, one in which $SU(5)_3$ is
already broken to $\gws$ and the standard matter transforms in
$5\oplus \bar{10}$ of $SU(5)_3$.
To simplify the discussion let us first focus on $SU(5)^3$.
In view of its  role we  treat $SU(5)_3$
as a weakly gauged factor, and neglect  possible effects coming
from its non-perturbative dynamics.

The space of $D$-flat directions of the theory is parametrized by 3
baryonic
invariants $u=\det(R)$, $v=\det(Q)$, $z=\det(\bQ)$ plus a meson $Y=RQ\bQ$.
A renormalizable superpotential consists just of the term
\beq
W=\lambda RQ\bQ=\lambda Y.
\label{tree5}
\eeq
This superpotential lifts most flat directions.
The classical equations of motion,
$\partial_R W=0$, $\partial_Q W=0$ and $\partial_{\bQ} W=0$, 
when contracted to form invariants give the constraints
\beq
uv=uz=vz=0 \quad\quad Y=0.
\label{flat5}
\eeq
Thus the space of classically flat directions 
 consists of three one-dimensional
baryonic branches $u,v,z$ with one common point at the origin.
Notice that this result holds also for the model in which one
gauge factor, say $SU(5)_3$, is not gauged\footnote {In that case the
space of $D$-flat directions consists of the same baryons
plus a meson matrix $\hat Y_A^B=(RQ\bQ)_A^B$, where $A,B$ are $SU(5)_3$ 
indices, supplemented by
the classical constraint $\det \hat Y=uvz$. Again the
superpotential is $W= \lambda \tr \hat Y$, and the condition
of $F$-flatness  leads to a vanishing meson matrix
$0=Q_B\partial_{Q_A} W=\lambda \hat Y_B^A$
and hence back to eq.~(\ref{flat5}). Thus even for a global
$SU(5)_3$ only the baryonic branches of moduli space survive.}.

The class of models to which this one belongs have been discussed
in ref. \cite{pts} where it was shown that in the presence of the
trilinear superpotential (and treating $SU(5)_3$ as weakly gauged)
they break supersymmetry at any finite point
in field space. Furthermore there is a sloping direction $z\to \infty$ along
which the vacuum energy becomes asymptotically zero.
However, the dynamics which  interests  us takes places
far away along the $u$ (or $v$ for that matter, since $SU(5)_1$ and  
$SU(5)_2$
can be interchanged) flat direction.  It is just a special case 
of the dynamics discussed in Section 2.
Along $u$, $R=X\times \one$, so that $u=X^5$. This breaks the factor 
$SU(5)_2\times SU(5)_3$ 
down to the diagonal $SU(5)_D$, which will be later identified with the
standard model gauge interactions.
Apart from $X$,  all the fields in $R, \,Q,\,\Qb$ get mass.
 Thus the $SU(5)_1$ gaugino condensation generates a superpotential
$W_{eff}=\Lambda_L^3=\lambda X \Lambda_1^2$.
A few comments are in order at this point. First of all,
this is now an exact result. There
are no corrections to it  coming from the $SU(5)_2$ dynamics. This is
not totally surprising, as that group is Higgsed. This result
can also be  established by studying
the global symmetries of the microscopic Lagrangian and using holomorphy.
In particular the baryon number $U(1)_Q$ with
charges $(0,0,1,-1,0,5)$
for $(\lambda, X, Q, \bQ, \Lambda_1^{10}, \Lambda_2^{10})$, rules out
any dependence of the low energy superpotential on $\Lambda_2$.

Now, let us analyze quantitatively the
 realistic scenario where $SU(5)_3\to \gwsp$.
To avoid confusion with gauge group suffixes we rename $SU(5)_2\to
SU(5)_B$ (balancing) and $SU(5)_1 \to SU(5)_S$ (strong).
The previous discussion is easily generalized.
Now $R\to R_2\oplus R_3=(\one,\five, \two\oplus\three)$ and
similarly $Q\to Q_2+Q_3$. The flat direction analysis is unchanged.
The vev of $R$ breaks  $SU(5)_B\times\gwsp\to \gws$, so that the
low energy gauge couplings are
\beq
{1\over g_i^2(X)}={1\over g_B^2(X)}+{1\over g_i'^2(X)}
\label{gaugedouble}
\eeq
where the notation is obvious.
Notice that at the scale $X$ the gauge couplings $g_i$ are modified
in an $SU(5)$ invariant manner, so that gauge unification is safe.
However, the coupling above $X$ will be somewhat bigger (like $\sqrt 2$).
This we find to be the major source of constraints for this model,
as we discuss in more detail below.

The $SU(5)_1$ nonperturbative dynamics goes along unmodified.
The field $R$ has however broken up into two
pieces $R_{2,3}$ whose wave function terms evolve differently
and which we indicate by $Z_{X_2}$ and $Z_{X_3}$.
Now, along the flat direction $R_3=X_3 \one_{3\times 3}$, $R_2=X_2
\one_{2\times 2}$ with $X_3/X_2=\sqrt{Z_{X_2}/Z_{X_3}}$ in
order to have vanishing $SU(5)_B$ D-terms. As before, we parametrize
the flat direction with $X^5=X_3^3X_2^2=\det R$.
The K\"ahler potential along $X$ is now given by
\beq
K(X,X^\dagger)=5 Z_{X_2}^{2\over 5}Z_{X_3}^{3\over 5} XX^\dagger
\label{twothree}
\eeq
so that the minimum condition is
\beq
5{\dot Z_X\over Z_X}=2{\dot Z_{X_2}\over Z_{X_2}}+3{\dot Z_{X_3}\over
 Z_{X_3}}=0
\label{min23}.
\eeq
The gauge coupling RGE are 
\beq
{d \over d \tau}({1\over g_B^2},{1\over g_S^2}, {1\over g_3'^2},
{1\over g_2'^2},{1\over g_1'^2})=(10,10,-2,-6,-{58\over 5}).
\label{gaugerg5}
\eeq
while for the Yukawa couplings we have
\bea
{d \ln (Z_X)\over d \tau}&=& {24\over 5} g_B^2+{3\over 5}g_2'^2
+{8\over 5} g_3'^2+{1\over 5}g_1'^2-2\labd-3\labt \\
{d \ln \labt\over d \tau}&=& 13\labt +2 \labd 
-{48\over 5}g_B^2 -{48\over 5}g_S^2-{16\over 3} g_3'^2-{4\over 15} g_1'^2 \\
{d \ln \labd\over d \tau}&=& 12\labt +2 \labt
-{48\over 5}g_B^2 -{48\over 5}g_S^2-3 g_2'^2-{3\over 5} g_1'^2. \\
\label{yukawarg5}
\eea
This model has no problems with $\ddot Z_X$ becoming positive.
This a consequence of $SU(5)_B$ being ``strongly'' asymptotically free.
In a situation like this, the minima in the $1/Z_X$ potential are
well pronounced. In fact they are so pronounced that the parameter 
space of this model
is strongly constrained by the avoidance of Landau poles in
$\labd,\labt$. The constraint is made even more acute by 
eq.~(\ref{gaugedouble}), which
implies that both $g_B^2$ and $g_i'^2$ have to be larger, in
practice by about a factor of 2, than the effective $g_i^2$.
This implies that the Yukawas that saturate $\dot Z_X=0$ have to be
rather large too, and their evolution above $X$ quickly reaches a Landau
pole. We find that the request of no Landau pole below $10^{17}$ GeV and
of perturbative gauge unification sets a lower bound on the scale X
\beq
X\gsim 10^{14}\gev
\label{boundx}
\eeq
For such an $X$ we find solutions with the following values of the
parameters: $g_B^2(X)\simeq 1-1.2$ and $\labd(M_G)\simeq \labt(M_G)\sim
1.8-3$ and $g_{GUT}^2<2$. This is still in the perturbative domain.
However, by lowering $X$ we immediately get large Yukawas at the GUT scale.
Notice that for $X\sim 10^{15}$
GeV, we start recovering the flavour problem because
of gravity contributions to
soft masses, thus weakening the motivation for gauge mediated supersymmetry
breaking.
So we can say that this model works in a fairly limited
but still non vanishing  range  $X\sim 10^{14}-10^{15}$ GeV.

Again we should comment on all additional contributions that may
affect the scalar potential. In this model, given the
small ratio $\Lambda/X\sim 10^{-5}$, the non-perturbative
corrections to the K\"ahler metric from the $SU(5)_S$ dynamics are
clearly negligible. But we are very close to the Planck scale.
The lowest operator that can affect the flatness is $\delta W=
{\rm det} X/M_{Pl}^2\sim X^5/M_{Pl}^2$. Then it is clear that for 
$X\sim 10^{14}$ GeV this correction must be absent, or be very
suppressed. If this is the case, flatness is preserved,
as the higher order terms like ${\rm det X}^2/M_{Pl}^7$ are very
suppressed. One last source of perturbation can arise from
 whatever mechanism relaxes to zero the cosmological constant
by providing an effectively  constant term $\sim \Lambda^2 M_{Pl}$
in $W$. This term creates in lowest order a linear potential
$\delta V\sim \Lambda^4 {\rm Re} X/M_{Pl}$. This leads to a shift $\delta X$
in the vev of $X$, which is 
\beq
{\delta X\over X}\sim (8 \pi^2)^2{X\over M_{Pl}}
\eeq
where the $(8\pi^2)^2$ arises from the expression for $m_X^2$, see eq. 
(\ref{xmass2}).
Then again, $X\sim 10^{14}$ GeV is still acceptable,
while $X\sim 10^{15}$ GeV starts being problematic. This bound
is amusingly similar to the one provided by the flavor problem.

In spite of its very constrained range of validity we think
this model is an attractive example of a very elegant
way of generating and communicating supersymmetry breaking.
Are there any special phenomenological signature of models
like this? The answer is in principle yes. Indeed in this
theory the messengers of supersymmetry breaking are not exhausted by
the chiral superfields $Q,\Qb$. There are also the massive
vector superfields in the coset $SU(5)_B\times \gwsp/\gws$. These
get a mass from $X$ and feel the effect of $F_X\not = 0$. A set of
these massive vectors transforms as the adjoint of $\gws$ and couples
directly to standard matter. We may call these heavy objects
the axial gluons, $W$'s and $B$. Indeed already at one loop
they affect the sparticle parameters. Since the field $X$ is the
only source of infrared cut-off, the form of the correction is
fixed by the 1-loop wave function renormalization, as for $K(X,X^\dagger)$
is Section 2
\beq
\delta \L_{soft}\sim \int d^4\theta {g_B^2\over 8\pi^2}
\ln(XX^\dagger/M_{Pl}^2) QQ^\dagger
\label{oneloopA}
\eeq
where the notation is the same as in eq.~(\ref{squarkop}). Notice the 
interesting
fact that this 1-loop correction generates only $A$ terms at order
$\alpha/4\pi$. The lowest contribution to the chiral preserving soft
masses is zero as $\partial _X\partial_{X^\dagger} \ln(XX^\dagger)=0$.
Chiral preserving mass squared  are generated at $\O(\alpha^2)$
by both the above 1-loop operator and by a host of new 2-loop effects.
In this way the correct scaling of gaugino and sfermion masses
is preserved. This cancellation of dangerous 1-loop effects is the same
as the one observed in ref. \cite{musol,DiNiSh} for the case of standard
matter directly coupled to the messengers via Yukawa couplings. We stress once
more that in the superfield language the cancellation simply follows
from the coincidence of the infrared regulator $X$
and the source of supersymmetry breaking. It does not work, for instance, when
the particles in the loop get mass from two independent spurion
superfields $X$ and $Y$ for which $X/Y$ is not a c-number. As it
was shown in ref. \cite{musol,DiNiSh} the cancellation works only
for the leading contribution in an expansion in powers of $(\Lambda/X)^2$.
The subleading terms correspond to higher derivative terms in the
effective action for $X$ and SM matter, which are not determined
by the RG log and for which the above argument does not apply. These
higher order effects are however unimportant for this model as $\Lambda/X
\ll 1$. A more detailed analysis is needed to determine the 
effects on the sparticle spectroscopy of the terms induced by the axial vector
superfields. It would be interesting to find experimentally important
differences. This we leave for further study. One sure thing is that
$A$ terms are now induced directly at the messenger scale and they are
of the same order as gaugino masses . This may have important implications for
the case in which the $\mu$ term in the Higgs sector is provided by
a singlet $S$ with couplings $W=S^3+SH_1H_2$. In conventional gauge
mediated models the smallness of the RG induced $A$ terms in
this sector makes the correct  breaking of the electroweak group 
more difficult. This is the case even when the messenger
scale is fairly high. It would be interesting to study the impact of
the new boundary conditions.

\section{Upper Limits on the Messenger Scale from Nucleosynthesis}

The messenger scale is $M$ is constrained by laboratory and  
astrophysical considerations. The first is the supersymmetric flavor problem,  
which is a primary motivation for gauge mediated theories.
It costrains the messenger scale to be less than about  
$10^{15}$ GeV~\cite{AHMRM}.

The purpose of this section is to introduce a stronger constraint  
coming from the requirement that the NLSP decays do not spoil the  
successful predictions of primordial nucleosynthesis.
These constraints will turn out to be fairly independent of the  
nature of the NLSP which is either a neutralino or the right handed  
stau depending on the number of messengers $N$ as well as the messenger  
mass $M$ as shown in fig.~3.
Let us first discuss the case when the NLSP is a neutralino, $\chi^0$.
The decays of $\chi^0$  are studied in detail in reference \cite{wells}.  
There it is shown that, in general, the dominant decay modes of $\chi^0$  
are
into a photon and gravitino and, if kinematically allowed, 
to a lesser extent into $Z^0$ and gravitino.
The decay rates are given by:
\beq
\Gamma (\chi^0\to \gamma {\tilde G} )\simeq \cos^2\theta_W 
\frac{m_{\chi^0}^5}{16\pi F^2}
\eeq
\beq
\Gamma (\chi^0\to Z{\tilde G} )\simeq \sin^2\theta_W 
\frac{m_{\chi^0}^5}{16\pi F^2}\left(1-\frac{m_Z^2}{m_{\chi^0}^2}\right)^4.
\eeq

The photon of the dominant decay mode can affect nucleosynthesis  
because it can break-up heavier nuclei into lighter ones  
\cite{earlier,starkman,ellis}. The ``damage''  that these decays can do  
to nucleosynthesis depends on the product of the mass of the decaying  
particle times its abundance relative to baryons (or photons):
\beq
d_\gamma \equiv m_{\chi^0}\frac{n_\chi}{n_B}.
\eeq
This quantity is somewhat model dependent, as it depends on the  
sparticle spectrum. For example, the lightest neutralino in typical  
gauge mediated theories is mostly a B-ino and its abundance relative to  
baryons depends on its self-annihilation rate and consequently on the  
slepton spectrum. This in turn depends on $N$ and $M$ and other features  
of the messenger sector \cite{multimess}. Nevertheless, the order of  
magnitude of $d_{\gamma}$ is easy to estimate provided that the typical  
mass scale of the light supersymmetric
particles is about 100 GeV, resulting in a self-annihilation 
cross section of order $2\pi (\alpha/m)^2\simeq 10^{-35}~{\rm 
cm}^2~m_{100}^{-2}$.
Here $m_{100}$ is the typical mass of light sparticles in units of  
100 GeV.
This yields a mass times abundance per unit entropy \cite{KolbTurner}:
\beq
mY_{\infty}\equiv m\frac{n_X}{s} \simeq 3\times 10^{-11}~{\rm GeV},
\eeq
corresponding to $d_{\gamma}\sim$ 0.3 $m_{100}$ GeV. This  
implies that  \cite{starkman,ellis} as long as the lifetime of the  
decaying particle is less than $10^7$ sec the damage that the photonic  
decays can do to nucleosythesis is negligible. They imply an upper  
limit of the scale of supersymmetry breaking 
$\sqrt{F} < 10^{10}$ GeV or equivalently
\beq
M<Nm_{100}^{3/2}~10^{15}~{\rm GeV},
\eeq
where $N$ is the number of messengers.
Therefore we see that the constraint from photodestruction  
of light elements is quantitatively similar
to that from the flavor problem.

%%%%%%%%%%%%%%%%
% FIGURE 3
%%%%%%%%%%%%%%%%
\begin{figure}[t]
%   \vspace{-4cm}
   \epsfysize= 8.5cm
   \epsfxsize=10.2cm
   \centerline{\epsffile{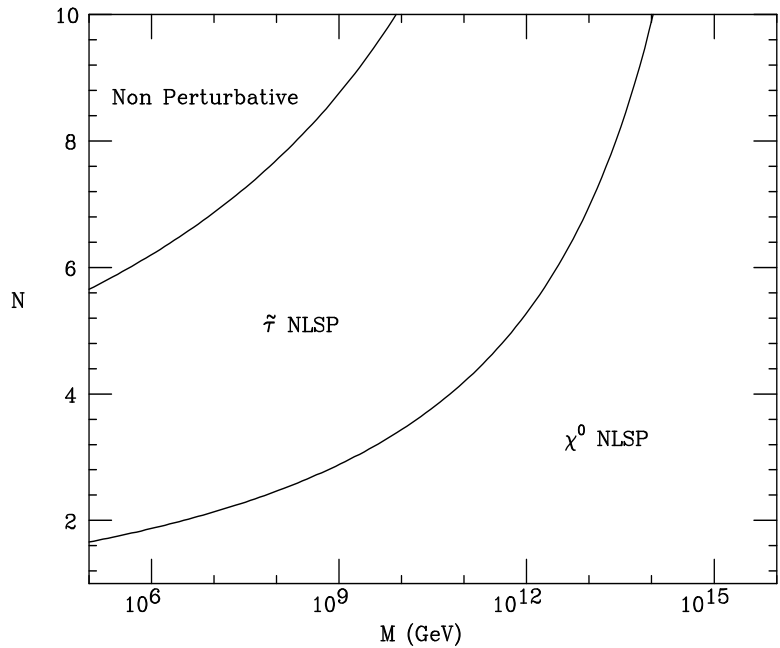}}
%   \vspace*{-15cm}
\caption{\label{plot} 
The regions in the $N$--$M$ parameter space where
the neutralino ($\chi^0$) or the right-handed stau (${\tilde \tau}$) are
the next-to-lightest supersymmetric particle (NLSP). Above the upper
curve the gauge coupling constants become non-perturbative before the
GUT scale.
}
\end{figure}
%%%%%%%%%%%%%%%%

A stronger constraint arises from the hadronic decays of 
$\chi^0$~\cite{starkman}. These  
occassionally lead to energetic nucleons which subsequently
thermalize by colliding with the ambient protons and $^4$He and in  
the process they create a ``hot line'' along which a one dimensional  
version of nucleosynthesis can take place, potentially resulting in  
overproduction of the light nuclei  D, $^3$He, $^6$Li, $^7$Li 
\cite{starkman}.  
D and $^3$He arise mostly from ``hadrodestruction'' -- an energetic  
nucleon breaks up an ambient $^4$He nucleus into $^3$He or D. $^7$Li and 
$^6$Li  
arise from ``hadrosynthesis'' in which  $^3$He or $T$ or $^4$He are  
sufficiently activated along the ``hot line'' to overcome their  
Coulomb barrier and synthesize $^6$Li or $^7$Li nuclei.
The analysis is fairly involved and the reader is referred to  
reference \cite{astrostarkman,starkman} for the details. The crucial  
quantity measuring the magnitude of the ``damage'' of hadronic decays  
on the primordial nuclear abundances is
\beq
d_B=d_\gamma r_B^*,
\eeq
where $r_B^*$ is essentially the hadronic branching ratio $r_B$ of the  
decaying particle  -- it is defined carefully in references  
\cite{astrostarkman,starkman}. The photon in the decay $\chi^0 \to \gamma  
{\tilde G}$ can convert to quarks with probability of order $\alpha$, 
and therefore $r_B$ can  
never be smaller than about  $10^{-2}$. The contribution to the  
hadronic branching ratio from the decay $\chi^0 \to Z  {\tilde G}$ is
\beq
r_B\sim 0.7~\tan^2\theta_W ~\left( 1-\frac{m_Z^2}{m_{\chi^0}^2}\right)^4,
\eeq
where 0.7 is the hadronic branching ratio of the $Z$.
This  is typically smaller than $10^{-2}$ as long as the $\chi^0$ 
mass is less than about
150 GeV because of phase space suppression. So from now on we take 
$r_B$ to be its minimal value $10^{-2}$; if it is larger then our  
arguments will get stronger. The value of the corresponding $d_B$ is  
$3\times 10^{-3}$ GeV.  From figures 5 of reference \cite{starkman} we see  
that even for such a small value of $d_B$ the hadronic showers lead to  
an extreme overproduction of $^7$Li, $^6$Li and D + $^3$He, 
if the lifetime of  
the decaying particle is longer than $10^4$ sec.
This shows that the lifetime of $\chi^0$ must be significantly smaller  
than $10^4$ sec.
How short the lifetime has to be before we are safe is not yet known  
to us. For lifetimes less than $10^4$ sec we are  
not aware of a complete analysis -- including the crucial hadronic  
jets -- of the effects of  late  decays on all the light element  
abundances, in particular $^6$Li and $^7$Li. One complication is that  
ordinary nucleosynthesis is occuring simultaneously with the decays.  
For this reasons it is not possible yet to quote  
a precise upper limit on the lifetime of a decaying particle from  
nucleosynthesis. A safe upper limit for the lifetime is about 1 sec, which  
marks the beginning of nucleosynthesis.
This leads to an upper limit of about $10^8$ GeV for the scale of  
supersymmetry breaking $\sqrt{F}$ or,
\beq
M<Nm_{100}^{3/2}~10^{11}~{\rm GeV}.
\label{limnucl}
\eeq
This is a safe upper limit on the messenger scale and is  
significantly stronger than the limit of $M< 10^{15}$ GeV coming from  
the flavor problem. However, it may be too strict and a detailed  
analysis may allow for somewhat longer  
lifetimes~\cite{progr}. 
Thankfully, the upper limit on $M$ scales as the square root  
of the lifetime, so that even if lifetimes of order 100 seconds were  
allowed they would only increase $M$ by an order of magnitude. This  
would still lead to an upper limit on M 
much stronger than the limit coming from  
the flavor problem. Since most nuclear abundances, in particular that  
of $^4$He, are in full bloom by the end of 180 seconds we expect that  
lifetimes longer than  180 seconds are excluded. The point is that  
the damaging processes of hadrodestruction and hadroproduction -- which  
lead to the overabundances of D+$^3$He and Lithium respectively -- rely  
mainly on the existence of ambient $^4$He which is completely formed by  
180 seconds. Furthermore, the universe is sufficiently diluted by 180  
seconds that a reduction of the excess abundance is unlikely.

 When the NLSP is a right-handed slepton the upper limit to the  
messenger scale should be even smaller than in eq.~(\ref{limnucl}).
The point is  
that the lightest right handed slepton is the right handed stau which  
decays into a tau and a Goldstino. Since the tau has a large hadronic  
branching ratio, the hadronic damage factor $d_B$ becomes 0.2  
GeV  which significantly exceeds that of $\chi^0$.

The upper limit in eq.~(\ref{limnucl})
has significant implications on model building.  
It disfavors theories in which the messenger scale results from the  
competition between renormalizable and non-renormalizable  
Planck-supressed operators, since these result in large messenger  
scales. It favors theories in which the messenger scale is the result  
of renormalizable competing forces, as is the case here. Even then, it  
excludes our most favored theory, namely $SU(5)^3$.

One, perhaps unappealing, way out of the constraint 
in eq.~(\ref{limnucl}) is
to postulate some degree of $R$-breaking which can result in a  
microscopic lifetime for the NLSP which is totally decoupled from 
supersymmetry breaking.

Finally, notice that the longest NLSP lifetime consistent with the  
flavor problem ($M<10^{15}$ GeV) is  $10^7$ sec. This implies that the  
NLSP decays cannot have an effect on the spectrum of the microwave and  
diffuse background radiation \cite{ellis,KolbTurner}.

It is interesting that the nucleosythesis constraint implies that the  
scale of supersymmetry breaking breaking has to be in the ``narrow''  
range between $10^4$ and $10^8$ GeV. To avoid the gravitino problem in  
the high end of this range -- between $10^6$ and $10^8$ GeV -- it is  
necessary that the reheating temperature of the universe be relatively  
low.

\section{Conclusions}

Gauge mediated supersymmetry breaking provides
a very attractive alternative to gravity mediated scenarios.
 In gauge mediation the only source
of flavor violation which is relevant in low-energy phenomenology
is given by the Yukawa matrices of standard quarks and leptons.
This leads to the supersymmetric generalization of the GIM mechanism
and to the distinctive experimental signature that no major
deviation from the SM is expected in flavor and CP violating
processes. This is consistent with
our present experimental information, and is the
main motivation for gauge mediation.  The gravity scenario,
on the other hand, has a generic difficulty with flavor violation. 
Nonetheless,
the models that can be made consistent with the present bounds 
({\it eg.} by using horizontal symmetries) 
still  predict that some major deviation
from the SM (like $\mu \to e \gamma$ or an
electric dipole moment of an elementary particle\cite{HKR})
will  sooner or later be discovered. Whether gravity or gauge interactions
mediate soft terms may  be answered by accurate low energy
experiments.

The breaking of supersymmetry and its mediation (via gauge interactions)
at scales much below $M_{Pl}$ should be well described by field theory,
with no active role played by quantum gravity. This motivates the study 
of theories where supersymmetry is broken by some non-perturbative
gauge dynamics. This allows for an attractive field theoretic  explanation
of the smallness of the weak scale itself, not just its quantum stability.
It is clear that the discovery of compelling field theory models
with dynamically driven and gauge mediated supersymmetry breaking
would give this scenario another  non-trivial boost over the gravity
mediated case. 

These facts are at the basis of numerous recent attempts to build
realistic theories wedding dynamical SUSY breaking to gauge mediation.
The main point of our investigation has been that many of the aesthetic
and phenomenological disadvantages of previous attempts are easily avoided
by relaxing the request that supersymmetry be broken in the true vacuum.
We have characterized a class of theories for which the vacuum where
we live is typically a false one, located on a supersymmetry
breaking potential energy plateau, far away from from other 
energetically more favourable minima. This last property can
be achieved quite naturally by means of Witten's inverse hierarchy
phenomenon and endows the false vacuum with a lifetime which
is much longer than the age of the Universe. This makes it in practice 
a stable vacuum. 

We have characterized a class of calculable theories with a 
fairly simple structure. The breaking of supersymmetry
is communicated to the MSSM sparticles in a minimal number of steps. 
A measure of this simplicity is the  fact that 
the Goldstino superfield itself gives a mass to the messengers.
This is not the case, for instance, in the standard scenario by
Dine {\it et al.} \cite{DiNeSh}, which contains an additional sector and where
the Goldstino is not directly coupled to the messengers.
Those models have no supersymmetric ground state, but this does not lead to
any clear advantage as the global minimum breaks color and $SU(2)_W$
at a high scale. Thus the messenger sector has to live 
in a local minimum. It seems that in order to have a phenomenologically 
acceptable
true minimum  considerable further complication is needed \cite{DaDoRa}. 
Notice that these minima are the analog of those
that exist in our models, see eq. (1), at $X=0$.
Indeed in gauge mediation with a large
messenger mass there can yet  be other deeper minima along
the flat directions $H_2=L_i$ \cite{komatsu} of the MSSM. 
This is because in gauge mediation the
soft mass  $m_{H_2}^2+m_{\tilde L}^2$ along this flat direction
goes negative very quickly due to the large stop contribution \cite{RaSa}.
So it seems that in one way or another gauge mediation brings
us the idea that we live in a false vacuum.

Attempts to build theories with just two sectors have flourished
in the last year. In particular the models proposed in Ref. \cite{PoTr}
,\cite{AHMRM} have a close resemblance to our $SU(5)^3$ example.
Those models are however phenomenologically non viable, as squarks
masses turn out to be negative. This is due to the presence of relatively 
``light'' matter supefields with positive and large supertrace.
 Since the potential is stabilized by $1/M_{Pl}$ operators, the scale 
 where soft terms (and the supertrace) are generated is fairly large
 ($\gsim 10^{15}$ GeV, barely compatible with the flavor problem), 
 the positive suprtrace leads to large and negative squark masses via
 two loop RG mixing. We stress that our $SU(5)^3$ model bypasses
 the problem by having no such light matter: everything gets a $\sim X$
 mass via just one Yukawa coupling. Other new mechanisms were explored
in Refs. \cite{LR,YS}, where the role of $X$ is played by a
composite gauge invariant field. Those models, however, involve
massive parameters in the superpotential, so they do not yet represent
a full microscopic theory.

In previous cases the messenger scale was either fairly low 
$\sim 10^{5}$ GeV, for renormalizable theories, or fairly high
$\gsim 10^{15}$, for non-renormalizable ones. In our scenario
we have a double dimensional transmutation, one for $F_X$ (gaugino
condensation) and one for $X$ (Coleman-Weinberg mechanism). By that we
can cover the continuum of theories where messengers have a
mass that varies from $10^{8-9}$ to $10^{14}$ GeV, and where $10^{6\div 7}
\gev \lsim F_X\lsim 10^{9}\gev$. Models that live in this
intermediate region can more easily avoid the gravitino problem
of usual renormalizable models \cite{murayamacosmology} and gravity 
has no observable flavor effect. Indeed we also pointed out that
models with a large messenger scale $\gsim 10^{12}$ GeV have serious
problems with standard Big Bang nucleosynthesis. This still keeps a
large slice of the continuum unconstrained. Models with $X>10^{12}$ GeV,
including  the otherwise very compelling $SU(5)^3$, are in need of
additional interactions which  allow the NLSP to decay in
less than a second. It seems that a possibility would be to
add a small source of $R$-parity violation, but with the aesthetic
and logical (remember flavor physics) problem that this addition entails.
Another totally different way out is to assume a period of GeV scale
inflation \cite{DiHa}, with GeV scale baryogenesis.

A possible direction for the future is to further enlarge the taxonomy
of theories with gauge mediated supersymmetry
breaking.  For instance, one hope is that with 
a large pool to choose from, a still unresolved problem like the origin
of $\mu$ may be solved in a natural way \cite{musol}. Indeed the boundary 
conditions
on soft terms plays a relevant role in this issue, making, for instance the
solution by the addition of a singlet $S$ more or less attractive.
A crucial aspect is the size of $A$ terms. We have seen that models
with heavy vector supefield messengers can lead to a rather
different boundary condition with respect to the minimal case, where
$A$ terms are generated by gaugino masses via RG evolution.

\section*{Acknowledgments}
We would like to thank R. Leigh, L. Randall, Y. Shadmi, M. Shifman, S. Thomas,
A. Vainshtein and  G. Veneziano for very useful and stimulating
conversations. We are especially indebted to A. Kusenko for
illuminating discussions on the bounce.

\section*{Note Added}
As this paper was about to be submitted a very interesting paper by 
H. Murayama,
hep-ph/9705271, appeared. In that paper a model similar
to our $SU(5)^3$ is discussed.
\newpage

\section*{Appendix}

In this brief Appendix we discuss the dynamics along
the flat direction $X$ as the field VEV becames of order $\Lambda$.
As an example we choose to illustrate the $SU(5)^3$ model, because
it is a very attractive one. Nonetheless, in the other
cases, the essence of the discussion is the same.

As already said the strong dynamics of $\gwsp$ can be
neglected.  The full effective superpotential of the confined
$SU(5)_1\times SU(5)_2$ theory is given by  \cite{pts}
\beq
W_{conf}= \lambda \tr \hat Y +A\left (\det(\hat Y)- uvz+ \Lambda_1^{10}u-
\Lambda_2^{10} v\right )
\label{confined}
\eeq
where $A$ is a Lagrange multiplier enforcing a quantum modified 
constraint \cite{quantumc}.
At least for $\lambda\ll 1$ we expect the above fields and $W_{conf}$ to 
describe the configurations of lowest energy.
Notice that the above superpotential admits no supersymmetric vacua
at finite field values. Indeed $F_Y=0$ requires $A,Y\not = 0$ after
which $F_z=0$ implies $uv=0$. This last requirement is inconsistent
with $F_u=F_v=0$, unless $z\to \infty$. Far away along $z$ all other baryons
and meson are massive, as it is already shown by the classical analysis,
and integrating them out yields the effective potential  
$W_{eff}=(\Lambda_1^2
\Lambda_2^2)/z^{1/5}$ \cite{pts}.
So supersymmetry is asymptotically recovered only
as $z\to \infty$: this theory has no global vacuum. We are however
interested in the $u$ branch and in what happens as $X$ is lowered
to be $\sim \Lambda_1$. In order to so, we first integrate out
the meson $Y$ and the field $A$. This is sensible as they are massive
over the entire 
moduli space. As a matter of fact the baryons $u,v,z$ are the only
fields that can become very light on some branch of the moduli space.
We thus obtain
\beq
W_{eff}=\lambda \left (uvz +u \Lambda_1^{10}-v\Lambda_2^{10}
 \right )^{1\over 5}
\label{bardestab}
\eeq
The above equations contains all the interesting information
regarding the dynamics as we move along $X=(u)^{1/5}$.
%Consider for the moment the limit $\lambda_2\ll \Lambda_1$,
%so that we can neglect the last term in brackets above, and focus
%on the mass matrix for the baryons $v,z$ along $u$.
For
$u\not = 0$ we can make the reparametrization $\tilde z=z-\Lambda_2^{10}/u$.
%For $X\gg \Lambda_1$
%a large supesymmetric mass drive $v, \tilde z$ towards the origin.
Then performing
the suitable rescaling of the baryons $v,\tilde z$
their mass matrix is roughly given by
\beq
\L_{mass}=(v^*, \tilde z^*)\left (\begin{array}{cc}\lambda^2 |X|^2 &  
\lambda   \Lambda_1^2\\
                             \lambda   \Lambda_1^2 &\lambda^2 |X|^2
                              \end{array} \right )
\left ( \begin{array}{c} v \\  \tilde z
                    \end{array}\right ).
\label{baryonmass}
\eeq
where the diagonal  terms come from $|F_{v,\tilde z}|^2$,
while the off diagonal from $|F_X|^2$.
At large $X$  the baryons are massive with zero vev,
so that by decoupling them we get back the linear $W_{eff}$.
   However,
when $X$ is decreased below some critical value $\sim \Lambda_1$
the baryons become tachionic and develop a vev. This indicates
that for $X\sim \Lambda_1/\lambda$ we enter the basin of attraction of the $z$
flat direction and then slide to infinity.
  The only point that is crucial for our discussion, however,
is that no instability develops for $X\gsim \Lambda_1/\lambda$, and 
thus to tunnel
to the  asymptotically supersymmetric vacuum the bounce configuration
has to interpolate between field values of order $X$ to values
of order $\Lambda_1/\lambda$. 

\def\ijmp#1#2#3{{\it Int. Jour. Mod. Phys. }{\bf #1~}(19#2)~#3}
\def\pl#1#2#3{{\it Phys. Lett. }{\bf B#1~}(19#2)~#3}
\def\zp#1#2#3{{\it Z. Phys. }{\bf C#1~}(19#2)~#3}
\def\prl#1#2#3{{\it Phys. Rev. Lett. }{\bf #1~}(19#2)~#3}
\def\rmp#1#2#3{{\it Rev. Mod. Phys. }{\bf #1~}(19#2)~#3}
\def\prep#1#2#3{{\it Phys. Rep. }{\bf #1~}(19#2)~#3}
\def\pr#1#2#3{{\it Phys. Rev. }{\bf D#1~}(19#2)~#3}
\def\np#1#2#3{{\it Nucl. Phys. }{\bf B#1~}(19#2)~#3}
\def\mpl#1#2#3{{\it Mod. Phys. Lett. }{\bf #1~}(19#2)~#3}
\def\arnps#1#2#3{{\it Annu. Rev. Nucl. Part. Sci. }{\bf #1~}(19#2)~#3}
\def\sjnp#1#2#3{{\it Sov. J. Nucl. Phys. }{\bf #1~}(19#2)~#3}
\def\jetp#1#2#3{{\it JETP Lett. }{\bf #1~}(19#2)~#3}
\def\app#1#2#3{{\it Acta Phys. Polon. }{\bf #1~}(19#2)~#3}
\def\rnc#1#2#3{{\it Riv. Nuovo Cim. }{\bf #1~}(19#2)~#3}
\def\ap#1#2#3{{\it Ann. Phys. }{\bf #1~}(19#2)~#3}
\def\ptp#1#2#3{{\it Prog. Theor. Phys. }{\bf #1~}(19#2)~#3}

\vskip .2in

\end{document}